\documentclass[twocolumn,amssymb,nobibnotes,showpacs,superscriptaddress,aps,pra,tightenlines]{revtex4-1}
\usepackage{color,graphicx,amsfonts,amssymb,amsmath,hyperref,footmisc,epsfig,fancyhdr}

\begin{document}

\title{Reinforcement Learning Enhancing Entanglement for Two-Photon-Driven Rabi Model}
\author{Tingting Li $^{\dag}$}
\affiliation{School of Science, Qingdao University of Technology, Shandong, China}
\author{Yiming Zhao $^{\dag}$}
\affiliation{School of Science, Qingdao University of Technology, Shandong, China}
\author{Yong Wang}
\affiliation{School of Science, Qingdao University of Technology, Shandong, China}
\author{Yanping Liu}
\affiliation{School of Science, Qingdao University of Technology, Shandong, China}
\author{Yazhuang Miao}
\affiliation{School of Science, Qingdao University of Technology, Shandong, China}
\author{Xiaolong Zhao$^*$}
\affiliation{School of Science, Qingdao University of Technology, Shandong, China}
\email{zhaoxiaolong@qut.edu.cn}
\date{\today}

\begin{abstract}
A control scheme is proposed that leverages reinforcement learning to enhance entanglement
by modulating the two-photon-driven amplitude in a Rabi model. The quantum phase diagram
versus the two-photon-driven amplitude and the coupling between the cavity field and the
atom in the Rabi model is indicated by the energy spectrum of the hybrid system, the witness
of entanglement, second order correlation, and negativity of Wigner function. From a dynamical
perspective, the behavior of entanglement can reflect the phase transition and the reinforcement
learning agent is employed to produce temporal sequences of control pulses to enhance the
entanglement in the presence of dissipation. The entanglement can be enhanced in different
parameter regimes and the control scheme exhibits robustness against dissipation. The
replaceability of the controlled system and the reinforcement learning module demonstrates the
generalization of this scheme. This research paves the way of positively enhancing quantum
resources in non-equilibrium systems.
\end{abstract}
%\pacs{03.65.Yz, 02.30.Yy, 03.67.-a}
\maketitle
\thispagestyle{fancy}
\lhead{}
\lfoot{\small{$^{\dag}$ These authors should be considered co-first authors.}}
\cfoot{}
\rfoot{}

\section{Introduction}
The interaction between a two-level system and an electromagnetic field, fully described
by the quantum Rabi model~\cite{PR51652,RPP691325} is a platform which has widespread
utility ranging from the study of fundamental physics \cite{Haroche2006} to the cutting
edge of quantum information~\cite{RMP91025005}. In the regime of cavity quantum
electrodynamics (QED), when the cavity-qubit coupling rate is strong enough to dominate
dissipation rates or reaches the domain of transition frequencies of the cavity or qubit,
the counter-rotating term cannot be ignored, and the system is best described by the
quantum Rabi model~\cite{RMP91025005,NRP1-19,PRL120-183601}. Once the coupling is strong enough,
a wide range of interesting phenomena occurs, including innovative quantum phase
transitions in the single-qubit-single-oscillator model when the frequency of the
oscillator is much lower than that of the qubit~\cite{PRA87-013826,PRL115180404},
generation of pure-cat-state~\cite{PRA81042311,PRA96043834}, strongly entangled states,
and novel steady-state squeezing~\cite{PRL131223604}. To date, in specially designed
architectures, the qubit-cavity coupling rate can reach the considerable fraction of
the cavity transition frequency~\cite{NP6772}. By inductively coupling a flux qubit and
an LC oscillator via Josephson junctions, ultrastrong-coupling regime with unconventional
transition spectra has been realized ~\cite{NP1344}. Quantum simulation in circuit QED
provides experimental access to the physics of the ultrastrong-and deep-strong coupling
regimes of an effective quantum Rabi model~\cite{PRX2021007,NC8779}. The strong coupling
limit of cavity QED has been demonstrated in a solid-state system by the vacuum Rabi
mode splitting and the corresponding avoided crossings~\cite{Nature431162}.

The periodically kicked pulsed pump field of a cavity in the presence of Kerr medium
acts as a parametric amplifier to investigate nonclassical behaviors of the mean photon
number~\cite{PRA444704}. In a cavity QED setup, the time-dependent two-photon parametric
driving can enhance the light-matter coupling and generate entanglement~\cite{PRL120-093601,PRL120-093602}.
Nonclassical cavity mode can be generated through a nonlinear optical parametric amplifier
with amplitude determined by pump field~\cite{PRL124073602}. Robust photonic cat states
can be generated by a two-photon process non-adiabatically~\cite{NPJQI318}. Similarly, the
generation and stabilization of cat state based on the interplay between Kerr nonlinearity
and single-mode squeezing has been demonstrated experimentally~\cite{Nature58413205}. A
review work provides a comprehensive perspective about quantum amplification and simulation
for strong and ultrastrong coupling of light and matter~\cite{PR1078-1}. We will combine the
time-dependent two-photon processes with a reinforcement learning module to prepare non-classical
quantum states in this work.

Machine-learning techniques are emerging as effective tools in exploring
physics~\cite{Murphy2012,Sutton2018,Murphy2018,PRX8031084}, and among them, reinforcement
learning (RL) has the potential to find optimal control fields to engineer system dynamics,
given that the evolution of these systems is governed by specific differential equations.
Time-varying control proposals have been employed to produce spin squeezing in a collective
spin systems~\cite{PRA103032601,AdP5362400056}. These works provide the fundamental model
to generate specific quantum states by reinforcement learning in this work. Reinforcement
learning can successfully steer the quantum Kapitza oscillator to a Floquet-engineered
stable inverted position amidst noise, with applications in quantum information and quantum
optics\cite{PRB98224305}. The application of RL to quantum control for a non-Markovian
open quantum system preparing the superposition of two excited states highlights the potential
of RL for quantum control in complex systems~\cite{PRA109013104}.

We first exhibit the phase diagram of a two-photon-driven quantum Rabi model to
delineate the parameter region to work on. Then we mainly propose an RL-based
control scheme to design the time-varying amplitudes of the two-photon-driven
strength to enhance the entanglement between the qubit and cavity mode. The
agent is trained to produce a sequence of 3-value square pulses to steer the
system to entangled states under the domain of a Lindblad master equation at
strong coupling regimes and the critical point for different initial states.

This paper is organized as follows: In Sec.\ref{Model}, we introduce the two-photon-driven
Rabi model. In Sec.\ref{PhaseDiagram}, the phase diagram versus the coupling strength and
two-photon-driven amplitudes in terms of several quantities are given. In Sec.\ref{EERL},
we present the procedure to enhance entanglement via reinforcement learning, including
reinforcement learning, dynamics and control results. In Sec.\ref{General}, we discuss the
generalization of this control scheme in terms of the controlled quantum system and the
replaceability of the reinforcement learning module. Finally, we conclude in Sec.\ref{CONC}.
\begin{figure}[htbp]
\includegraphics[width=\linewidth]{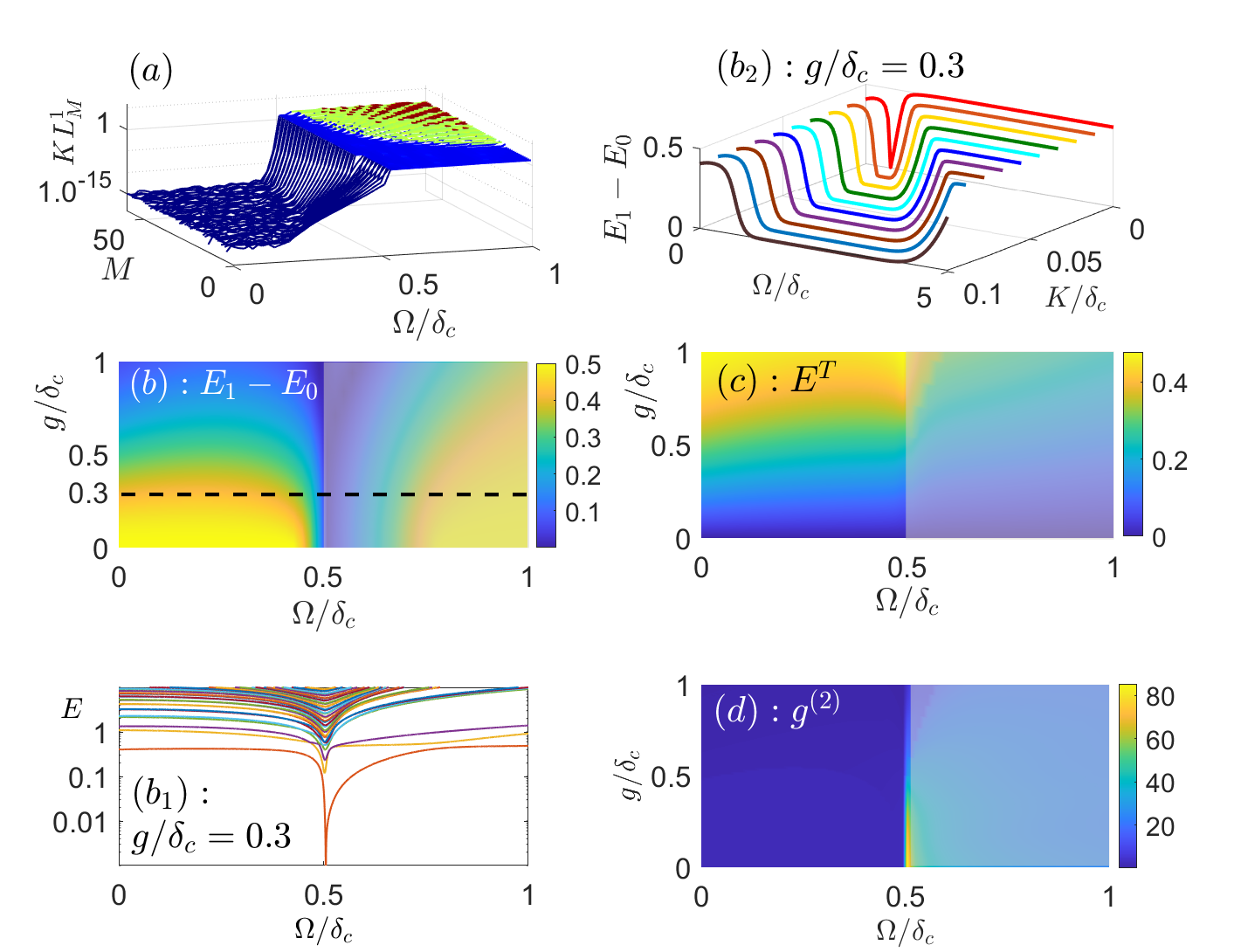}
\caption{$(a)$ $KL_M^1$ versus the truncated dimension $M$ and $\Omega/\delta_c$
when $g/\delta_c$ =0.1 (logarithmic scale used). $(b)$ The energy gap between the
first excited state $E_1$ and the ground state $E_0$ versus $g/\delta_c$ and
$\Omega/\delta_c$, $M=60$ and $K/\delta_c = 0$ hereafter. $(b_1)$ The energy
spectrum $E-E_0$ versus $\Omega/\delta_c$ when $g/\delta_c =$ 0.3. $(b_2)$ The energy
gap $E_1-E_0$ versus $\Omega/\delta_c$ and $K/\delta_c$ when $g/\delta_c$ = 0.3.
$(c)$ The partial transposed criteria for entanglement of the ground state $\rho_0$,
denoted by $E^T$ versus $g/\delta_c$ and $\Omega/\delta_c$. $(d)$ Equal time
second order correlation versus $g/\delta_c$ and $\Omega/\delta_c$. Areas where
numerical simulations might fail are overlaid with a white semi-transparent shadow.}
\label{FigPhaseDiagram}
\end{figure}
\section{The model}
\label{Model}
Quantum Rabi model~\cite{PR51652,RPP691325} describes the interaction between a
single-mode bosonic field (such as a cavity mode) and a qubit (generic two-state
system). It is described by the Hamiltonian ($\hbar=1$ hereafter)
\begin{equation}
\begin{aligned}
\hat{H} =&\delta_c \hat{a}^{\dag}\hat{a}+\frac{\delta_q}{2} \hat{\sigma}_z+g(\hat{a}+\hat{a}^{\dag})(\hat{\sigma}_-+\hat{\sigma}_+)\\
&+\Omega (\hat{a}^2+\hat{a}^{\dag 2})+K\hat{a}^{\dag 2}\hat{a}^2,
\end{aligned}\label{Eq:Hamiltonian}
\end{equation}
where $\hat{a}^{\dagger}$($\hat{a})$ is the creation (annihilation) operator for
the bosonic cavity mode and $\hat{\sigma}_{z}$ is the Pauli $z$-basis operator
with commute relation $[\hat{\sigma}_+,\hat{\sigma}_-] = \hat{\sigma}_z$.
$\delta_{c(q)}=\omega_{c(q)}-\omega_p$, where $\omega_{c}$, $\omega_{q}$
and $\omega_{p}$ are the frequencies of the cavity, qubit and parametrical driving,
respectively. The parameter $g$ is the coupling strength between the two subsystems.
The two-photon-driven term $\Omega(\hat{a}^2+\hat{a}^{\dag 2})$ can be implemented
by Kerr nonlinear
Hamiltonian~\cite{PRL120-093601,PRL120-093602,PRA444704,NPJQI318,PRL124073602,Nature58413205}.

In the regime of weak coupling, through the rotating wave approximation, the quantum
Rabi model can be effectively described by the Jaynes-Cummings model~\cite{JPB38S551},
which has been investigated widely in circuit QED system \cite{PRX2021007}, cavity QED
system~\cite{RMP73565}, and trapped ions \cite{PRL118073001,PRL128160504}. When
the coupling strength approaches or surpasses the frequency scales of the cavity mode
and the qubit, the rotating wave approximation no longer holds and remarkable phenomena
emerge~\cite{NRP1-19,RMP91025005}, like virtual excitations, asymmetric anticrossing of
the polariton excitation branches~\cite{PRA74033811}, and avoided-level crossings not
included in the Jaynes-Cummings model~\cite{NP6772}, novel thermodynamic character~\cite{PRE98012131},
nonlinear optical phenomena~\cite{PRA95063849}, superradiant phase~\cite{PRA87-013826,PRL115180404}
and nonclassical states include squeezed states, Schr$\ddot{o}$dinger-cat states, and entangled states~\cite{PRA81042311}.
We will examine the phase diagram with respect to the two-photon-driven strength and the coupling
rate between the subsystems. Then the dynamical behavior to enhance the entanglement by
reinforcement learning method is investigated.

\section{The Phase Diagram}\label{PhaseDiagram}
\begin{figure}[htbp]
\includegraphics[width=\linewidth]{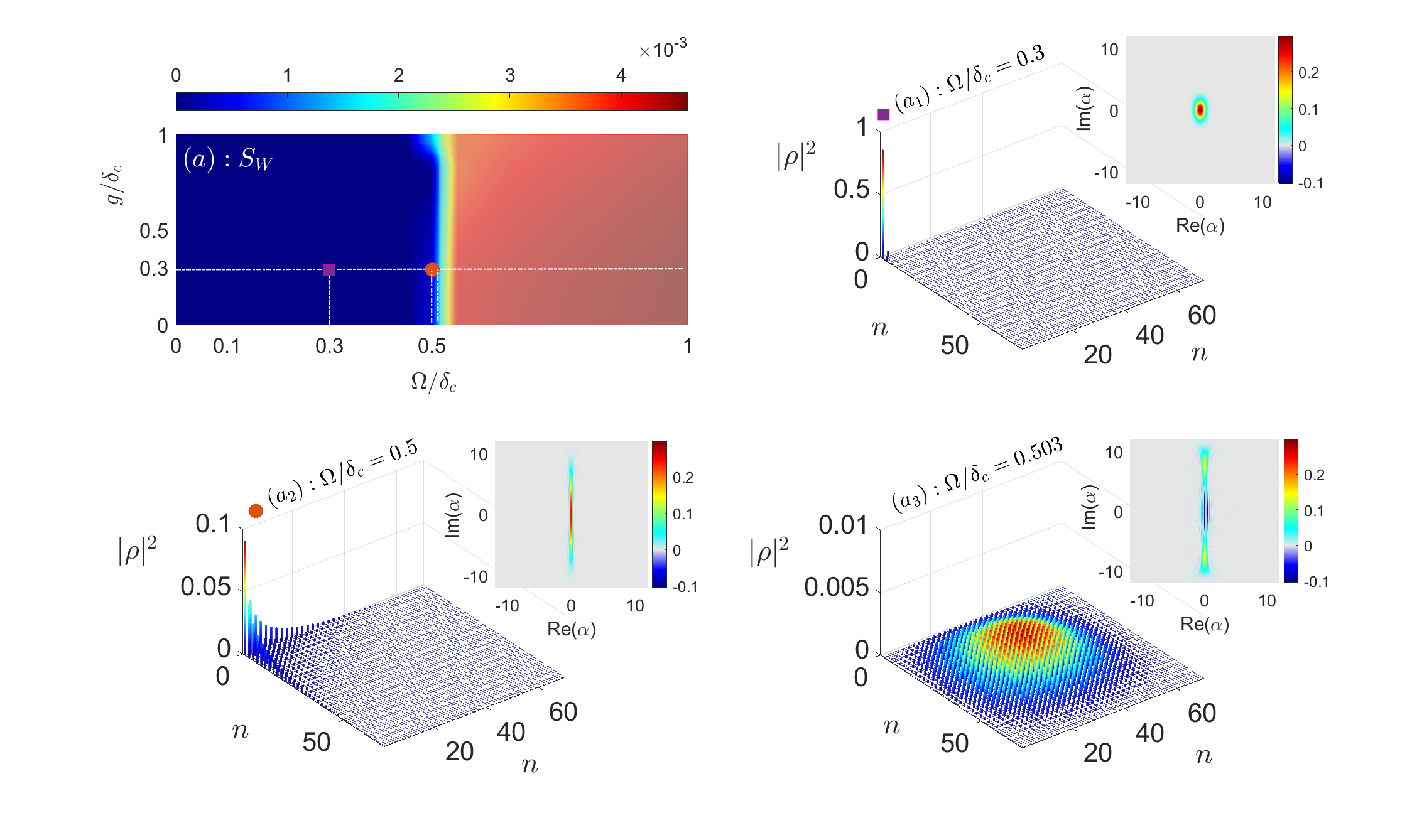}%
\caption{$(a)$ The average of evenly distributed 10000 sampling points of the Wigner
function versus $g/\delta_c$ and $\Omega/\delta_c$ for the cavity ground state.
$(a_1)$ - $(a_3)$ The distribution of $|\rho|^2$ for the cavity ground states when the
coupling strength $g/\delta_c = 0.3$ and $\Omega/\delta_c = 0.1, 0.5, 0.503$ with the
corresponding Wigner functions, respectively.}
~\label{WignerFun}
\end{figure}

If a system has different phases, it possesses distinguishable properties in different
phases. We conjecture there is a phase transition in the two-photon-driven Rabi model,
which may be examined by the behavior of several quantities versus the strength of the
parameter-driven amplitude $\Omega/\delta_c$ and coupling strength $g/\delta_c$ in the
Hamiltonian~(\ref{Eq:Hamiltonian}). In addition to the energy gap between the ground
state and the first excited state, the partial transpose criteria for entanglement,
second correlation, and negative value of Wigner function are checked in Fig.~\ref{FigPhaseDiagram}
and Fig.~\ref{WignerFun}.

The number of excitations in the cavity field in the ground state may increase dramatically
as the parameters in the Hamiltonian vary. In order to ensure the accuracy of numerical
simulations, a higher truncation dimension is required. To ensure the accuracy of numerical
simulations, inspired by the concept of Kullback-Leibler divergence (relative entropy)~\cite{AMS22-79},
we define a quantity denoted as $KL_M^q$ :
\begin{equation}
KL_M^q = \rho_{M+q} |\log\rho_{M+q}-\log\rho_{M}|,
\label{KL}
\end{equation}
where $\rho_{M}$ is the density matrix for the ground state of dimension-$M$ eigenspace
and the integer $q$ represents the dimension difference between the two density matrices.
0-padding is used at the higher truncation margin of the density matrix $\rho_{M}$ to
compensate for the truncation difference between $\rho_{M}$ and $\rho_{M+q}$. When calculating
the logarithm of the matrix elements, 0 matrix elements will be replaced by 1. This metric
is used to quantify the difference between $\rho_{M}$ and $\rho_{M+q}$. A smaller value of $KL_M^q$
indicates a larger similarity between the two states. As shown in Fig.~\ref{FigPhaseDiagram}
$(a)$ with $K/\delta_c=0$, $KL_M^1$ diverges when $\Omega/\delta_c > 0.5$. This divergence
behavior of $KL_M^1$ occurs with the continuous change of $\Omega/\delta_c$, so
\begin{equation}
\Omega_c = \frac{\delta_c}{2}.
\label{Eq:CCS}
\end{equation}
is a critical point for a phase transition. Thus we will investigate the dynamics in the
regime $\Omega/\delta_c \leq 0.5$.

This phase transition can also be indicated by the energy gap between the first excited
state and the ground state of the Hamiltonian~(\ref{Eq:Hamiltonian}) as shown in
Fig.~\ref{FigPhaseDiagram} $(b)$. As shown in $(b_1)$, the energy spectrum exhibits a
tendency towards degeneracy at $\Omega_c$. These results coincide with the that in
Fig.~\ref{FigPhaseDiagram} $(a)$. The Kerr nonlinearity $K$ in the Hamiltonian~(\ref{Eq:Hamiltonian})
can be seen as a kind of photon-dependent photon number
shift~\cite{PRL120-093601,PRL120-093602,PRA444704,NPJQI318,PRL124073602,Nature58413205}.
The influence of this Kerr nonlinearity on $E_1-E_0$ can be seen in
Fig.~\ref{FigPhaseDiagram}($b_2$). With increasing of $K/\delta_c$, the region of energy gap
$E_1-E_0=0$ extends. In this work, we focus on the dynamical character while enhancing
entanglement by machine learning with $K/\delta_c=0$.

The partial transpose criterion involves analyzing the eigenvalues of the partially
transposed density matrix of the composite system, and it offers a sufficient (but
not necessary) criterion for witnessing entanglement in bipartite quantum
systems~\cite{PRL771413,PRA65032314}. The appearance of negative eigenvalues of the
partially transposed density matrix is definitely a signature of entanglement. The
absolute of the negative eigenvalues of the partially transposed density matrix
(denoted by $\lambda_i$) are summarized to act as the entanglement witness:
\begin{equation}
\label{ET}
E^T=\sum_{\lambda_i<0}|\lambda_{i}|.
\end{equation}
As shown in Fig.~\ref{FigPhaseDiagram}($c$), the phase diagram indicated by $E^T$
coincides with the results in Fig.~\ref{FigPhaseDiagram} $(a)$ and $(b)$.

In the strong and ultra-strong coupling regimes, to exclude unphysical quadrature
measurements, an effective approach for describing the system involves solving the
eigen-equations of the Hamiltonian (\ref{Eq:Hamiltonian}) to construct the dressed
space. Then the cavity-field operators are redefined with positive and negative
frequencies.
$\hat X^{+}=\sum_{j,k>j}X_{jk}|j\rangle \langle k|$ and $\hat X^{-}=(\hat X^{+})^{\dagger}$,
with $X_{jk}\equiv \langle j|\hat a^{\dagger}+\hat a|k\rangle$, in the dressed
eigen-basis $|j\rangle$, $|k\rangle$ of the Hamiltonian (\ref{Eq:Hamiltonian}) with
eigen-values $\omega_j$ and $\omega_k$, respectively~\cite{PRA88063829,PRA92063830}.
 In the limit of weak coupling, these operators coincide with $\hat{a}$ and $\hat{a}^\dag$,
 respectively. And similar operators can be defined for $\hat \sigma_-$ and $\hat \sigma_+$
 \cite{PRA88063829,PRA92063830}. The master equation when applied to describe the dynamics,
would be modified by these dressed operators.

The strong correlation of emitted photons can now be decided by the equal time
second order correlation defined as
\begin{equation}
\label{FanoF}
g^{(2)}=\frac{\langle \hat{X}^{+2}\hat{X}^{2}\rangle}{\langle \hat{X}^{+}\hat{X} \rangle^2}.
\end{equation}
The behavior of $g^{(2)}$ versus $\Omega/\delta_c$ and $g/\delta_c$ is shown in
Fig.~\ref{FigPhaseDiagram}($d$) coincides the phase transition in Fig.~\ref{FigPhaseDiagram}($a$).

To gain insight into the phase transition further, we examine the Wigner function of the
ground states versus system parameters. The appearance of negative values in the Wigner
function is a sufficient but not necessary condition for characterizing the non-classicality
of a quantum state. We use the average of the negative values of the Wigner function as
\begin{equation}
\label{Wsum}
S_W=\frac{1}{M}\sum_n ^M W_{<0}
\end{equation}
to show the non-classicality of the cavity state, where $W_{<0}$ are the negative values
of $M$ evenly distributed sampling point of the Wigner function. $S_W$ versus $\Omega/\delta_c$
and $g/\delta_c$ is shown in Fig.~\ref{WignerFun}$(a)$. The sudden change of  $S_W$ confirms
the phase transition in Eq.~(\ref{Eq:CCS}). The Wigner functions of the ground state in different
phases are shown in Fig.~\ref{WignerFun}$(a_1)$, $(a_2)$ and $(a_3)$ with the density matrices
$|\rho|^2$ of the cavity ground states. Upon $\Omega/\delta_c > \Omega_c$, $S_W$ changes abruptly
with negative texture in Wigner function emerges. Meanwhile, the distribution of $|\rho|^2$ versus
the index of the basis $n$, changes obviously.
\section{Enhance Entanglement by Reinforcement Learning}\label{EERL}
\begin{figure}
\includegraphics*[width=\linewidth]{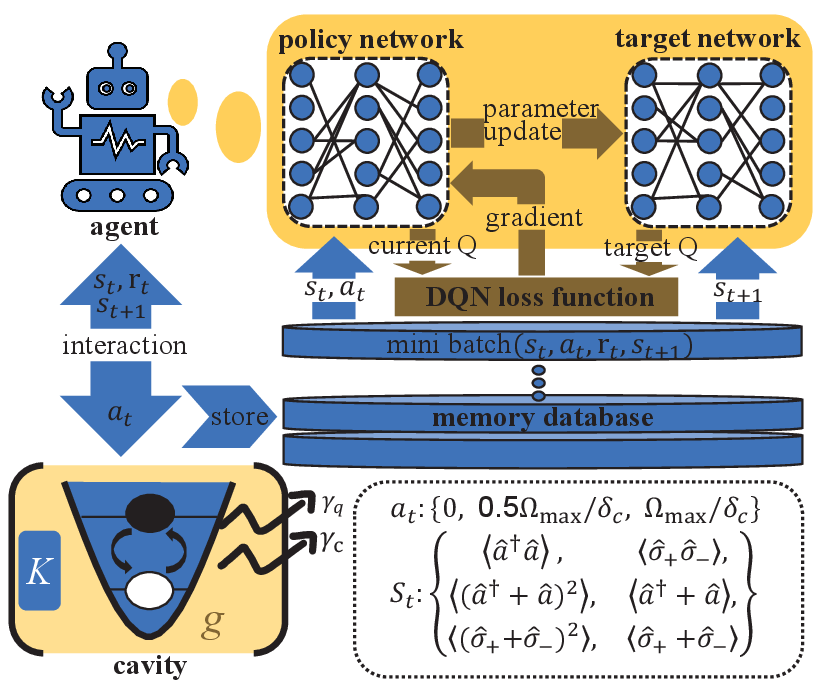}
\caption{The workflow of using reinforcement learning agent to enhance the entanglement for the two-photon-driven Rabi model. The hyper parameters for the DQN agent are set as follows: Batchsize = 128. The discount factor is set to 0.99. The starting value of $\varepsilon$ for the greedy strategy is set to 0.9 and the final value of $\varepsilon$ is 0.05, with exponential decay 1000. The update rate of the target network is set to 0.005. The learning rate of the AdamW optimizer is set to $1 \times 10^{-4}$. The neural network architecture used by the DQN agent is defined as follows: The network takes as input the observable properties of the quantum system is set to 6. Hidden Layers include layer1: fully connected layer has input size = 128 and output size = 128 with ReLU activation function. layer2: fully connected layer has input size = 128 and output size = 512, also with ReLU activation function. layer3: fully connected layer with input size = 512 and output size = 256, also with ReLU activation function. Output Layer: fully connected layer with input size = 256 and output size = 3 correspond to the possible actions. The output layer directly returns the Q-values without using an activation function.}
\label{RLPro}
\end{figure}

Taking a cue from control strategies by Lyapunov control and machine learning~\cite{PLA425127874,AdP5362400056},
we propose a scheme harnessing an RL agent to design temporal control field to enhance the
entanglement and check the influence of system parameters. Drawing inspiration from time-dependent
parameter process~\cite{PRL120-093601,PRL120-093602,PRA444704,NPJQI318,PRL124073602,Nature58413205},
in this control proposal, the Hamiltonian for the controlled quantum system reads:
\begin{equation}
\begin{aligned}
\hat{H} = &\delta_c \hat{a}^{\dag}\hat{a}+\frac{\delta_q}{2} \hat{\sigma}_z+g(\hat{a}+\hat{a}^{\dag})(\hat{\sigma}_-+\hat{\sigma}_+)\\
&+\Omega(t) (\hat{a}^2+\hat{a}^{\dag 2}),
\end{aligned}\label{CHamilt}
\end{equation}
here $\Omega(t) = \Omega_{max} f(t)$ with $f(t)$ is the control sequence designed by the RL agent,
and $\Omega_{max}$ is the amplitude. In this control process, $[\hat{a}^{\dag}\hat{a}+\hat{\sigma}_z+(\hat{a}+\hat{a}^{\dag})(\hat{\sigma}_-+\hat{\sigma}_+),\hat{a}^2+\hat{a}^{\dag 2}]\neq0$
ensures the effectiveness of the control fields. The workflow for this control scheme is illustrated
in Fig.~\ref{RLPro}. The control field $\Omega(t)$ acts as the sequence of the actions and the increase
of $E^T$ is positively related with the reward function in the reinforcement learning module.
\subsection{Reinforcement Learning}
\label{RLandGA}
Starting with no prior knowledge about the system under control, RL uses the trial-and-error paradigm
to iteratively learn the mapping between actions and states that maximizes the accumulated reward over
time. An appropriate rewards-evaluation rule which favors particular state-action mappings can
enhance the control performance. Decision-making executed by the agent entails selecting actions
$a_t=\pi(s_t)$ that manipulate the system changing from a state (not referring to quantum states,
but a set of expectations of operators of the controlled system) $s_t$ to $s_{t+1}$, with $\pi$
denoting the policy being learned~\cite{Murphy2012,Sutton2018}.

$Q$-learning operates through a $Q$ function that represents the expected total future reward for a
given policy $\pi$~\cite{Nature518529,PNAS38716}. The optimal policy $\pi^*$ with the maximized $Q$
function satisfies Bellman equation which encapsulates the principle of optimality for decision-making
over time. Deep $Q$-learning network (DQN) has been proposed to approximate this function~\cite{Nature521436,CS2500611,PMLR387395,PMLR19281937,arXiv151106581}.

\subsection{Enhancing entanglement by machine-designed pulses}
In DQN, the reward function is designed with the aim of achieving the control objective. In our
reinforcement learning proposal, the entanglement witness is employed to establish the reward
function. In this work, the reward function we employ is
\begin{equation}
\label{RewardF}
R_t = 10\langle\Delta E^T_t > 0\rangle-\langle\Delta E^T_t \leq 0\rangle,
\end{equation}
where $\langle\bullet\rangle$ is the sign function of $\bullet$ and
$\Delta E^T_t = E^T_{(t+1)}-E^T_{(t)}$. Here $E^T_{(t)}$ is the partial transpose criterion
for entanglement at discrete sampling time $t$. This means the agent gets 10 points of reward
when the entanglement increases and deducts 1 point when the entanglement decreases. This is
consistent with the core principles of reinforcement learning, the agent is incentivized to
provide actions to steer the system towards states with higher entanglement.

In this work, the RL agent selects those actions $a_t$ at sampling time $t$, contingent
upon the observations $S_t = \{\langle \hat{a}^{\dag}\hat{a}\rangle,
\langle \hat{\sigma}_{+}\hat{\sigma_{-}}\rangle, \langle (\hat{a}^{\dag}+\hat{a})^2\rangle,
\langle \hat{a}^{\dag}+\hat{a}\rangle,\langle (\hat{\sigma}_++\hat{\sigma}_-)^2\rangle,
\langle \hat{\sigma}_++\hat{\sigma}_-\rangle\}$, thereby orchestrating a sequence of actions
consist of $a_t \in \{0, 0.5\Omega_{max}/\delta_c, \Omega_{max}/\delta_c\}$ aimed at maximizing
cumulative rewards and minimizing penalties based on $R_t$. During the training, the observations
are fed into the neural network, while output neurons provide the probability of choosing which
action at the next iterative training step. A reward is dispensed subsequently to each step to
evaluate the decision-making policy. After one epoch, the controlled system is re-initialized
to the same initial state and the next epoch starts to train the agent continuously based on
the trained neural network. The state evolves according to the environment dynamics, which is
deterministic~(\ref{Master-eq}). The learned policy $\pi^*$ is represented by
the arrangement of the pulse sequences.

It is necessary to take measures to optimize the performance of reinforcement learning. One
is introducing the replay structure, which stores past experiences during training to
improve both learning efficiency and stability. To enhance the robustness of the reinforcement
learning agent against large errors resulting from the use of linear functions, the Huber
loss is employed~\cite{AMS3573}. This loss function not only avoids underestimating or
overestimating the influence of small errors, but also helps to mitigate the exploding
gradient problem during the training of deep neural networks, which benefits convergence
compared to other loss functions~\cite{AMS3573}. The neural network parameters are updated
by the algorithm AdamW~\cite{60RfadamW} which applies the decay directly to the weights.
It is beneficial in suppressing overfitting. Yet, Huber loss and AdamW can be replaced by
other loss functions and optimization algorithms~\cite{Murphy2012,Sutton2018}. The inherent
randomness in the reinforcement learning algorithm introduces a probability that enables
the agent to explore and identify the most effective sequence of actions.

It is essential to emphasize that the proposed strategy operates within a closed-loop
framework during simulation, while the actual implementation follows an open-loop control
scheme. Once the control sequence for $\Omega(t)$ in the Hamiltonian (\ref{Eq:Hamiltonian})
is obtained through simulation, the same control field is applied in an open-loop
process to avoid quantum collapse caused by direct observation on the system.
\subsection{Entanglement dynamics in open conditions}
\label{Dynamics}
Usually, it is hard to avoid decoherence in a quantum system due to its interaction with
the environment. The effect of such decoherence should be taken into account in this
control scheme to ensure reliability. There are two kinds of decoherence channels:
dissipation of the cavity field and the qubit, which ruin the entanglement. The system
would evolve under the domain of the Lindblad master equation with the application of
the time-dependent two-photon-driven term $\Omega(t)=\Omega_{max}f(t)$ in the Hamiltonian
(\ref{Eq:Hamiltonian}) designed using reinforcement learning.

In the strong and ultrastrong coupling regimes, the realistic scenarios with the environment
temperature $T$, is described by the master equation
\begin{equation}
\label{Master-eq}
\dot\rho(t) = i [\rho(t), \hat{H}] + \mathcal{L}_{\hat{a}}\rho(t) + \mathcal{L}_{\hat{\sigma}_{-}}\rho(t),
\end{equation}
where $\mathcal{L}_{\hat{a}}$ and $\mathcal{L}_{\hat{\sigma}_{-}}$ are Liouvillian superoperators
describing the decoherence of the cavity field and qubit~\cite{PRA84043832}.
They read
$ \mathcal{L}_{\hat{\chi}}\rho(t) = \sum_{j,k>j}\Gamma^{j k}_{\hat{\chi}}\bar{n}(\Delta_{k j},T)\mathcal{D}[|k \rangle \langle j|]\rho(t) + \sum_{j,k>j}\Gamma^{j k}_{\hat{\chi}}(1 + \bar{n}(\Delta_{k j},T))\mathcal{D}[|j \rangle \langle k|]\rho(t)$ for $\hat{\chi} = \hat{a} (\hat{\sigma}_{-})$
represents the cavity field (qubit), $\mathcal{D}[\mathcal{O}]\rho(t)$ =
$\frac{1}{2} (2 \mathcal{O}\rho(t)\mathcal{O}^{\dagger}-\rho(t) \mathcal{O}^{\dagger} \mathcal{O} - \mathcal{O}^{\dagger} \mathcal{O}\rho(t))$.
The relaxation coefficients $\Gamma^{j k}_{\hat{\chi}} = 2\pi d(\Delta_{k j}) \alpha^{2}_{\hat{\chi}}(\Delta_{k j})| C^{\hat{\chi}}_{j k}|^2$
with $d(\Delta_{k j})$ being the spectral density of the baths, $\alpha_{\hat{\chi}}(\Delta_{k j})$ denoting
the system-bath coupling strength, while $\Delta_{k j} = \omega_{k} - \omega_{j}$, and
$C_{j k}^{\hat{\chi}} = -i \langle j |(\hat{\chi} - \hat{\chi}^{\dagger})| k \rangle$ .
$\bar{n}(\Delta_{k j},T)$ ($\bar{n}$ for short) is the thermal population at frequency $\Delta_{k j}$ with
environment temperature $T$. When considering a cavity coupling to the momentum quadrature of a field in
one-dimension waveguides, the spectral density $d(\Delta_{k j})$ is constant and
$\alpha_{\hat{\chi}}^{2}(\Delta_{k j}) \propto \Delta_{k j}$. Then the relaxation coefficients read
$\Gamma^{j k}_{\hat{\chi}} = \gamma_{\hat{\chi}} \,(\Delta_{k j}/\omega_{0}) \, |C^{\hat{\chi}}_{j k}|^2$
where $\gamma_{\hat{\chi}}$ is the damping rate. These assumptions can be realized in circuit-QED.
The influence of dephasing and Lamb shifts in current experiments can be deemed negligible as they
do not exert significant influence~\cite{PRL118073001,PRL128160504}.
\begin{figure}
\includegraphics*[width=\linewidth]{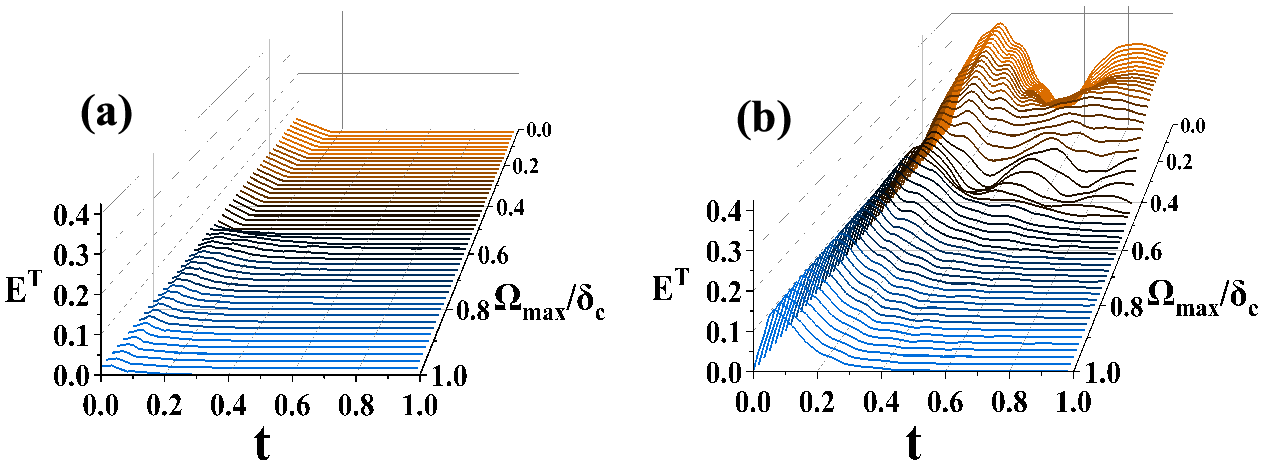}
\caption{Evolution of the entanglement indicated by partial transposed criterion $E^T$ versus the amplitude
of the two-photon-driven parameter $\Omega_{max}$ for the initial state $\rho(0) = \rho_0$ is the ground
state of the Hamiltonian~(\ref{Eq:Hamiltonian}) in (a), and $\rho(0)$ = $|\alpha=1,\downarrow\rangle\langle \alpha=1,\downarrow|$
in (b). $\delta_q/\delta_c$ =1, $g/\delta_c$ = 0.3, $K/\delta_c$= 0, and the dissipation rates
$\gamma_c/ \delta_{c}$ = $\gamma_q/ \delta_c$ = 0.01.}
\label{EntgPhase}
\end{figure}

In this study, we put forward a conjecture that the dynamics of the entanglement can reflect the phase
transition for different initial states. We first check the dynamics of the entanglement versus the
constant two-photon-driven strength when $g/\delta_c = 0.3$ in Fig.~\ref{EntgPhase}. As shown in
Fig.~\ref{EntgPhase} ($a$), the strength and duration time of the entanglement for the ground state
decreases as $\Omega/\delta_c$ increases. This character coincides with the results in
Fig.~\ref{FigPhaseDiagram} ($c$). This means the two-photon process has a negative effect
on the entanglement. While the initial state $\rho(0)$ = $|\alpha=1, \downarrow\rangle\langle \alpha=1,\downarrow|$,
where $|\alpha=1\rangle$ is the coherent state with amplitude 1, the entanglement exhibits more
fluctuations versus time when $\Omega_{max}/\delta_c$ is below the phase critical point $\Omega_c$.
Comparing Fig.~\ref{EntgPhase} ($a$) and ($b$), the entanglement increases more obviously
when the initial state is the non-entangled product state than the entangled ground state.
\begin{figure}
	\includegraphics*[width=\linewidth]{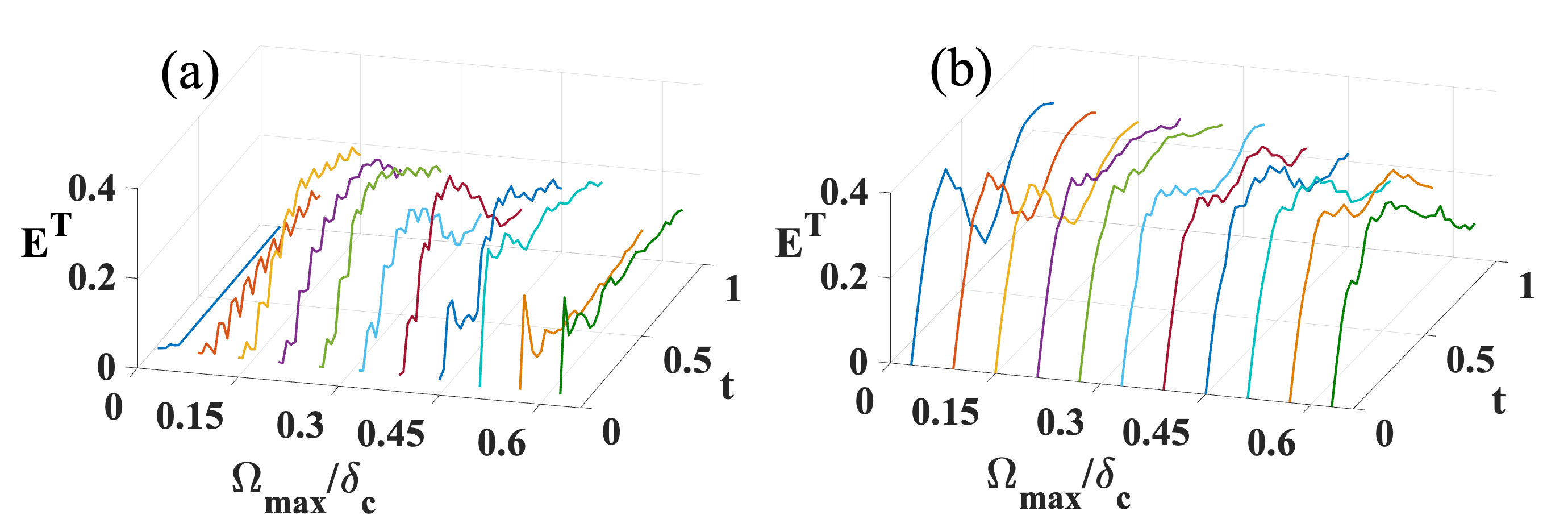}
	\caption{(a) Evolution of entanglement witness $E^T$ versus $\Omega_{max}/\delta_c$ for
$\rho(0) = \rho_0$ (the ground state of the Hamiltonian (\ref{Eq:Hamiltonian})), and (b)
$\rho(0)$ = $|\alpha=1, \downarrow\rangle\langle\alpha=1, \downarrow|$. $\delta_q/\delta_c$ = 1,
$g/\delta_c$ = 0.3, $K/\delta_c$ = 0, and dissipation rates $\gamma_{\hat{a}}/\delta_c$ =
$\gamma_{ \hat{\sigma}_{-}}/ \delta_c$ = 0.01.}
\label{CAR}
\end{figure}

\subsection{Entanglement under control}
\label{ConRes}
We conjecture the amplitude of the control field may have effect on the enhancing effect.
For the initial states $\rho(0) = \rho_0$, namely, the ground state of the Hamiltonian
(\ref{Eq:Hamiltonian}) and $\rho(0)$ = $|\alpha=1, \downarrow\rangle\langle\alpha=1, \downarrow|$,
we check the influence of the maximum amplitude $\Omega_{max}/\delta_c$ on the optimized
entanglement-enhancing case in Fig.~\ref{CAR}. This provides a reference for selecting the value
of $\Omega_{max}/\delta_c$. In the following research, we mainly focus on $\Omega_{max}/\delta_c$,
which is a moderate choice with respect to the system parameters.

\begin{figure}
	\includegraphics*[width=\linewidth]{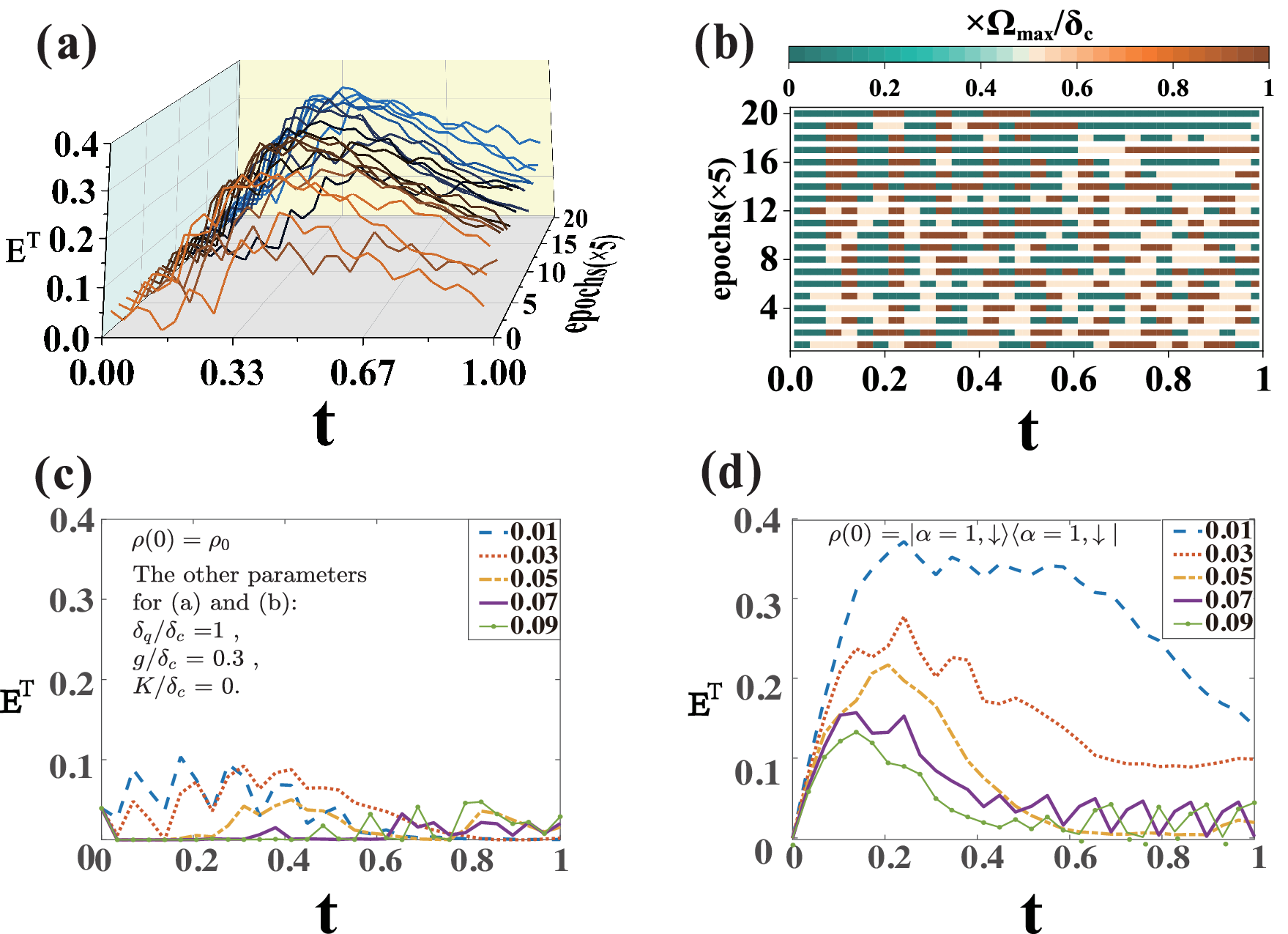}
	\caption{(a) The dynamics of the entanglement indicated by $E^T$ versus the training epoches.
(b) The corresponding square control pulses $\Omega(t)$ ($\Omega_{max}/\delta_c$ = 0.3) from
the top view, with values shown in the color bar. $\rho(0) = \rho_0$ is the ground state of the
Hamiltonian~(\ref{Eq:Hamiltonian}), $\delta_q/\delta_c$ = 1, $g/\delta_c$ = 0.3, $K/\delta_c$ = 0,
and dissipation rates $\gamma_{\hat{a}}/\delta_c$ = $\gamma_{ \hat{\sigma}_{-}}/ \delta_c$ = 0.01.
(c) and (d) Evolution of entanglement witness $E^T$ for different dissipations as shown
in the legend: $\gamma_c/ \delta_{c}$ = $\gamma_q/ \delta_c$ = 0.01, 0.03, 0.05, 0.07, 0.09, for
$\rho(0) = \rho_0$, the ground state of the Hamiltonian (\ref{Eq:Hamiltonian}) and $\rho(0)$ =
$|\alpha=1, \downarrow\rangle\langle\alpha=1, \downarrow|$, respectively.}
\label{F5exampleETG}
\end{figure}

\begin{figure}
	\includegraphics[width=\linewidth]{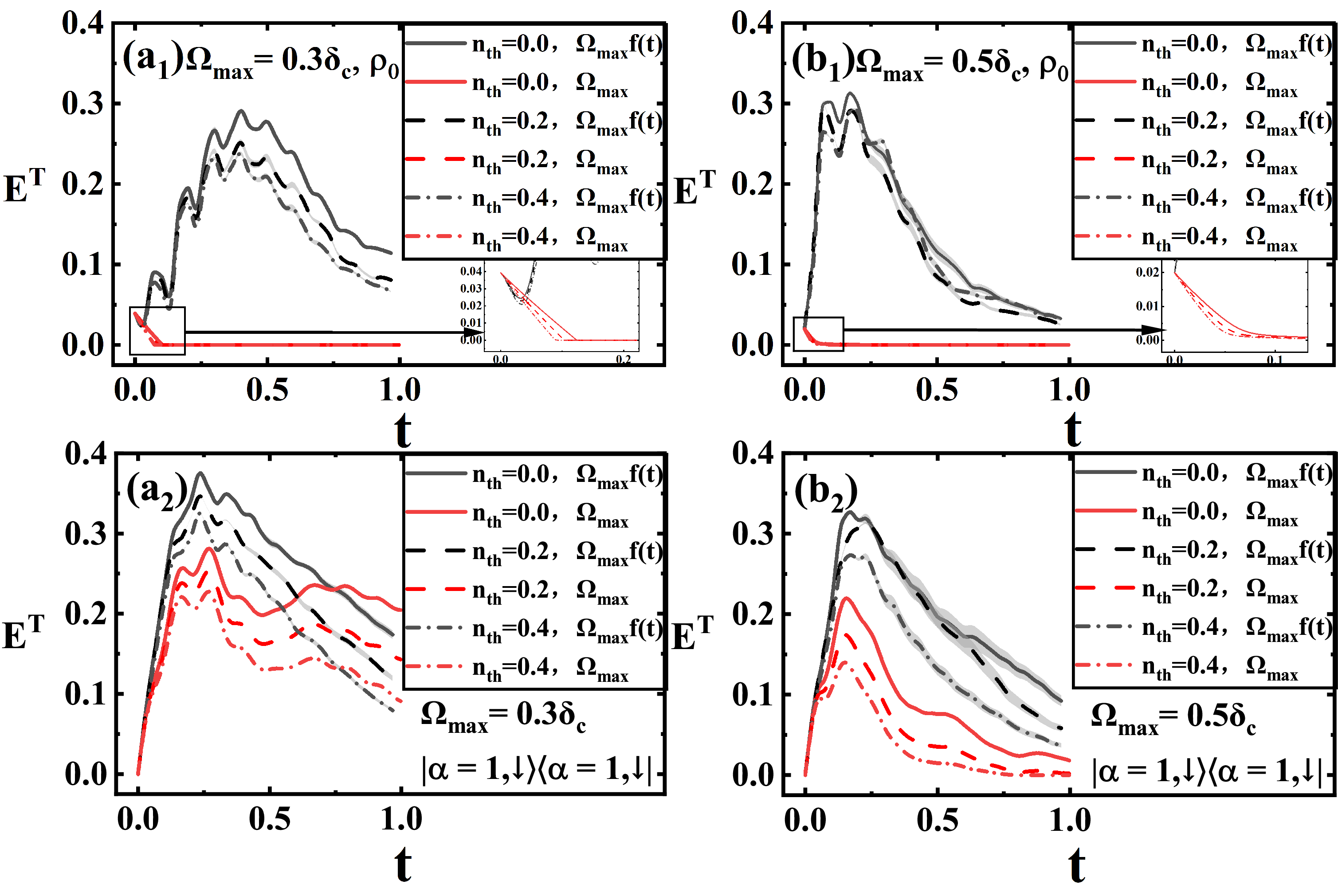}
	\caption{The average evolutions of 20 entanglement criterion $E^T$ with error zones for
different two-photon-driven amplitudes and initial states.
$(a_1)$ $\Omega_{max} = 0.3\delta_c$ and $\rho(0) = \rho_0$, the ground state of the
Hamiltonian (\ref{Eq:Hamiltonian}). $(a_2)$ $\Omega_{max} = 0.3/\delta_c$ and
$\rho(0)$ = $|\alpha=1, \downarrow\rangle\langle\alpha=1, \downarrow|$. $(b_1)$
$\Omega_{max} = 0.5\delta_c$ and $\rho(0) = \rho_0$ is the ground state of the Hamiltonian
(\ref{Eq:Hamiltonian}). $(b_2)$ $\Omega_{max} = 0.5\delta_c$ and
$\rho(0)$ = $|\alpha=1, \downarrow\rangle\langle\alpha=1, \downarrow|$. The other parameters
are the same to those in Fig.~\ref{F5exampleETG}.}
\label{F6_entgmode_for_RL}
\end{figure}

We check the dynamics of the entanglement witness $E^T$ under the control versus different system
parameters and initial states. The time interval [0,~1] is partitioned into 30 segments of control
square pulses with amplitudes picked in $\{0,~0.5\Omega_{max}/\delta_c,~\Omega_{max}/\delta_c\}$
determined by the RL agent. The RL agent designs the arrangement of the pulses under the objective
of increasing the entanglement. Each round of the training consists of 100 epochs. As an example,
one round of the control process to find the sequence of square pulses is shown in Fig.~\ref{F5exampleETG}.
As shown in Fig.~\ref{F5exampleETG}($a$), along with the training, the RL agent can find numerous
sequences of square pulses all with increased entanglement. The corresponding control sequences
are shown in Fig.~\ref{F5exampleETG}($b$). This control method can be regarded as a kind of
combination of bang-bang control~\cite{PRA582733,PRA86022321} and reinforcement learning. From a
statistical view, it can be intuitively observed that as learning progresses, the probability of
$\Omega_{max}/\delta_c = 0$ decreases in the distribution of the control pulses. Yet, the
negative effect of dissipation on control performance can be observed in Fig.~\ref{F5exampleETG}($c$)
and ($d$). With the same other parameters, the stronger the dissipation, the worse the effect of RL
enhanced entanglement. These solutions corroborate the efficacy of the control pulses designed by
the RL agent to enhance the generation of entanglement.
Zero temperature was assumed in the previous results. To investigate the robustness of the protocol
in reality, it is natural to consider the influence of temperature on the control results. Since
$\bar{n} = \frac{1}{\exp^{\hbar\omega/k_BT}-1}$, the average number of thermal excitations in the
reservoir $\bar{n}$ is positively correlated to the temperature $T$ and the strength of the
decoherence according to the master equation (\ref{Master-eq}). To examine the influence of
temperature on the control effect at different system parameters and initial states, we show the
dynamics of $E^T$ at different $\bar{n}$s in Fig.~\ref{F6_entgmode_for_RL}. From a statistical
perspective, larger $\bar{n}$ is detrimental to the enhancement of entanglement, whether or not
the proactive control by RL is applied to the system. From a holistic perspective, the application
of the control field designed by the RL agent has a positive effect on enhancing the entanglement.
By comparing $(a_1)$ with $(a_2)$ ($\Omega_{max}/\delta_c = 0.3$), or $(b_1)$ with $(b_2)$
($\Omega_{max}/\delta_c = 0.5$), the entanglement dynamics become similar when the control is
applied. And the entanglement can be enhanced more obviously by the control when the initial
state is the entangled ground state. By comparing $(a_1)$ with $(b_1)$ , or $(a_2)$ with $(b_2)$,
larger two-photon-driven strength is detrimental to the production of entanglement. At the
critical point, i.e., $\Omega_{max}/\delta_c = 0.5$, the negative impact of temperature on
entanglement enhancement is not significant enough. In this control scheme, it is reasonable
that one should choose the best control sequence among the solutions.

\section{Generalization Capability of The Scheme}\label{General}
The RL agent finds solutions to optimization problems through a trial-error mechanism
based on statistical principles. In a quantum-well waveguide system, the light-matter
interaction can be turned from weak to ultrastrong within less than one cycle of
light~\cite{Nature458178}, which provides the probability to implement this control
scheme. The tunability of the single-photon strong-coupling strength may be the candidate
to produce quantum squeezing~\cite{PRL114093602}. In a word, any other systems under
certain mapping rules (such as differential equations) with tunable parameters can be
considered as the controlled system in this proposal. Not only the controlled system
but also the RL agent can be replaced by other RL modules that perform similar functions.
For example, several alternatives have been identified and evaluated, including the
State-Action-Reward-State-Action algorithm~\cite{CS2500611}, Deep Deterministic Policy
Gradient~\cite{PMLR387395}, Asynchronous Advantage Actor-Critic~\cite{PMLR19281937},
and the Dueling Network~\cite{arXiv151106581} and so on.

\section{Conclusion}
\label{CONC}
In a Rabi model driven by a two-photon parameter, besides the energy gap between the first
excited state and the ground state, the entanglement criterion, equal-time second-order
correlation, and negativity of the Wigner function in the ground state also reflect the
phase transition. The entanglement dynamics exhibits a significant change at the critical
boundary. The reinforcement learning agent was employed to design a suite of square pulses
to modulate the time-varying two-photon parameter strength, thereby enhancing the entanglement
dynamically in an open environment. The entanglement is still enhanced, despite stronger
dissipation impairs more severely to the control performance. The versatility of the control
scheme is demonstrated by the potential to substitute the reinforcement learning agent with
other optimization modules and replace the controlled system with other systems that have
tunable parameters. This work provides the avenue to study non-equilibrium control problems
using reinforcement learning agents.

\section{Acknowledgements}
X. L. Zhao thanks discussions with Wenzhao Zhang, Natural Science Foundation of Shandong
Province, China, No.ZR2020QA078, No.ZR2023MD064, ZR2022QA110, and National Natural Science
Foundation of China, No.12005110.

\end{document}